# Technologies to support self-determination for people with intellectual disability and ASD

Laronze Florian[1*][0000-0001-5450-9732], Audrey Landuran[1] and N'Kaoua Bernard[1]

[1] INSERM, Bordeaux Population Health Research Center, UMR 1219, University of Bordeaux, Bordeaux, France;
*corresponding author

`florian.laronze@u-bordeaux.fr, audrey.landuran@gmail.com, bernard.nkaoua@u-bordeaux.fr`

**Abstract.** This article focuses on the concept of self-determination and the design and validation of digital tools intended to promote the self-determination of vulnerable people. Self-determination is an essential skill for carrying out daily activities. But in certain situations, and for certain populations, self-determination is lacking, which leads to the inability to live an independent life and in favorable conditions of well-being and health. In recent years, self-determination enhancing technologies have been developed and used to promote independent living among people with self-determination disorders. We will illustrate the main digital tools to support self-determination developed for two populations of people suffering from self-determination disorders: people with an intellectual disability and people with an autism spectrum disorder. The ability of these digital assistants to improve the comfort of life of these people will also be presented and discussed.

## 1    Self-determination

Wehmeyer (1992) developed a functional model of self-determination which is currently the most used (Landuran & N'Kaoua, 2018), particularly in the field of disability.

In this model, Wehmeyer (1999) proposes a definition of self-determination (completed a few years later) as "the set of skills and attitudes required in a person, allowing him to act directly on his life by making choices free, not influenced by undue external agents, with a view to maintaining or improving one's quality of life" (Wehmeyer, 2005). The principle of causal agent is a central element in this model. It implies that the person acts intentionally in order to bring about an effect to accomplish a specific objective or to bring about or create a desired change (Wehmeyer et al., 2011). Self-determination does not reflect a total absence of influence and interference, but rather consists of making choices and taking decisions without excessive and undue interference (Lachapelle et al., 2005).

According to this model, self-determined behavior refers to actions that are identified by four essential characteristics: (a) autonomy: the person is able to indicate preferences, make choices and initiate actions without external influence, (b) self-regulation: ability to regulate their behavior according to the characteristics of the environment and its behavioral repertoire in order to respond to a task, (c) psychological empowerment: the person has a feeling of control over their actions and the consequences of their actions on the environment, (d) self-realization: the person has a knowledge of herself (strengths, weaknesses, etc.) which allows her to adjust her choices and decisions according to her characteristics (Wehmeyer, 1999).

## 2  Intellectual disability and self-determination

Intellectual disability (ID) affects about 1% of the population, and involves problems that affect functioning in two areas: intellectual functioning (such as learning, problem solving, judgement) and adaptive functioning (activities of daily life such independent living). Additionally, these difficulties must have appeared early in the developmental period. The ID is extremely heterogeneous on the clinical and etiological levels and is characterized, on the cognitive level, by sensory and motor deficiencies (Bruni, 2006) in short-term, working and episodic memories (Jarrold, Baddeley, 2001), language (Rondal, 1995), and executive functioning (Lanfranchi et al., 2010). Related to these cognitive difficulties, many studies have shown that people with ID encounter difficulties in their daily life (Van Gameren-Oosterom et al., 2013), and are less self-determined than their non-disabled peers (Wehmeyer & Little, 2013).

However, numerous studies have shown that being self-determined improves quality of life (Lachapelle et al., 2005), independent living or even academic and professional success (Wehmeyer & Schwartz, 1997). Supporting the self-determination of people with ID is therefore a major issue. It is a question of allowing these people to live as much as possible according to their choices, their wishes, their desires, their preferences and their aspirations, without the handicap being a factor of exclusion (Nirje, 1972).

### 2.1  Technologies to support self-determination and Intellectual Disability

The rise of digital technology has made it possible to open up new and extremely promising lines of investigation. Indeed, in recent years, digital technologies, and in particular self-determination support technologies, have shown extremely positive results in people with ID, particularly with regard to social inclusion or community participation (Lachapelle et al., 2013).

These technologies now cover many areas of daily life. For example, communication support technologies offer expression aids to translate non-verbal communication behaviors (pressing an image, symbol, etc.) into synthesized or digitized verbal messages

(Soto et al., 1993), computer-assisted reading devices (Sorrell et al., 2007) or even Enhanced and Alternative Communication devices that complement or replace the production language, for example, through the use of pictograms (Kagohara et al., 2013). Assistive technologies for social interactions can assist people in recognizing simple emotions (fear, joy, sadness, anger), solving interpersonal problems or even conversation skills (Wert & Neisworth, 2003). Learning support technologies enable the acquisition of skills such as mathematics (Bouck et al., 2009), reading (Haro et al., 2012), cognitive skills (Brandão et al., 2010), motor skills or even the capacity for sensory integration (Wuang et al., 2011) which corresponds to the ability to interpret and organize effectively the information captured by the senses.

Other technologies aim to support daily activities, such as: managing one's budget (Mechling, 2008), using an automated banking machine (Alberto et al., 2005), paying for purchases (Ayres, et al., 2006), running errands (Bramlett et al., 2011), doing laundry, washing dishes (Cannella-Malone et al., 2011), setting the table (Lancioni et al., 2000), cleaning up (Wu et al., 2016), setting the table (Ayres et al., 2010), putting away groceries (Cannella-Malone et al., 2006), using the bus (Davies et al., 2010), learning new routes (Brown et al., 2011), making navigation decisions autonomously in order to reach unfamiliar places (McMahon et al., 2015) or even time management (Ruiz et al., 2009).

We should also note the support technologies for professional activities allowing the acquisition of professional skills (Allen et al., 2012), work-related social skills (Gilson, Carter, 2016), or making appropriate decisions when performing a professional task (Davies et al., 2003).

Finally, support technologies for leisure activities allow, for example, the learning of complex game sequences (D'Ateno et al., 2003), the use of and access to entertainment videos (Kagohara, 2011), transferring music to an MP3 (Lachapelle et al., 2013), community inclusion in the library (Taber-Doughty et al., 2008), access to digital documents and the Internet (Stock et al., 2006).

Positive repercussions have been noted in people using these technologies, such as an increase in self-confidence, sense of self-efficacy, motivation, self-esteem, identity development, self-determination or quality of life (Näslund & Gardelli, 2013).

While assistive technologies for self-determination have been proposed to help carry out many activities of daily living, no digital assistance has been proposed to help people with disabilities project themselves into the future, set goals and to develop life projects. It is in this context that we designed the digital assistant "It's my life! I choose it" for help with decision-making and the development of a life plan, for people with disabilities (Landuran & N'Kaoua, 2021).

## 2.2 Design and validation of a digital assistant for decision-making and development of a life project

If the notion of self-determination emphasizes the importance of freely consented choices, the notion of life project brings a dimension of projection into the future, of capacities to imagine one's own life according to one's desires and expectations. In the field of disability, the life project occupies a central place in supporting people (Nair, 2003). Many studies have shown that the development of a life plan, life goals or even personal goals has positive consequences, in particular on health, well-being, personal development or even quality of life (Cross & Markus, 1991).

The digital assistant "It's my life! I choose it" has 4 main sections. The first "Choosing, what is it? consists in defining and proposing exercises on the notion of choice (what is a choice, what we can choose, what we cannot choose, etc.). In the second part "My life", the user is invited to answer questions about his past and his present, in order to help him project himself into the future. The third part "What is important to me" makes it possible to reflect on the notion of value (what are the important values, what are the values that we do not share, etc.). Finally, in the fourth part "My project", the participants are encouraged to reflect on what they wish to do later in many areas such as family life, leisure, education, professional life, etc. The person is invited to consider the projects they wish to carry out in different areas (housing, work, affective life and leisure), to classify them according to their interests, and to try to define the needs necessary to achieve them (Fig. 1).

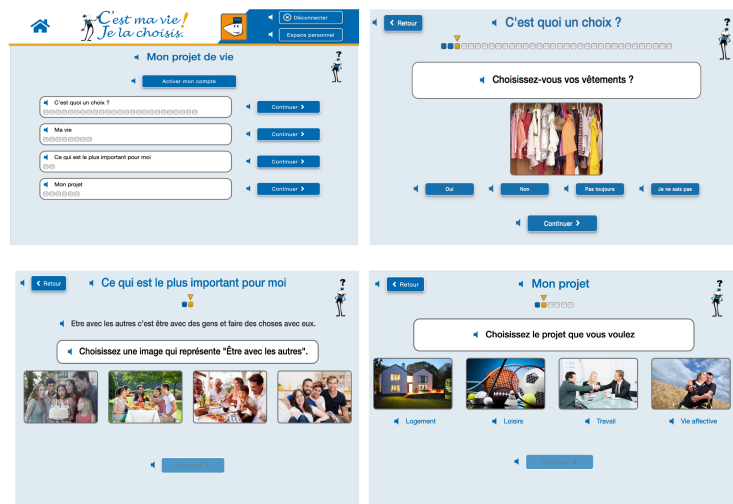

**Fig. 1.** Examples of digital assistant pages.

The repercussions of the use of the digital assistant on self-esteem, the level of worry, self-determination, psychological well-being and the formulation of the life project were evaluated with adults presenting a Trisomy 21.

The analysis of the results shows that the assistant makes it possible to: 1) improve the richness of the life project of adults with trisomy 21 in the different areas of the life project; 2) increase the feeling of self-determination, more particularly in the field of interpersonal relations; 3) increase the feeling of well-being in the dimensions of autonomy, positive relationships with others, meaning in life and self-acceptance; 4) increase self-esteem and 5) decrease worry (Fig. 2.).

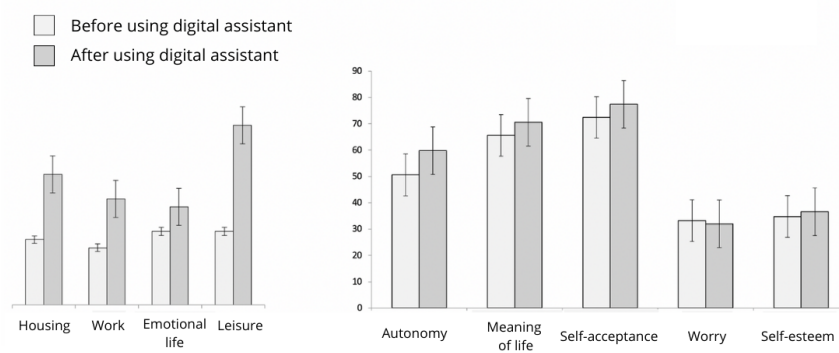

**Fig. 2.** To the left: Wealth of the life project in the different areas, before and after using the digital assistant. To the right: Perception of psychological well-being, level of worry and self-esteem, before and after using the digital assistant.

Another disabling activity limitation linked to intellectual disability is the inability, for some young adults, to access residential autonomy, which forces them to live with their parents or in specialized accommodation. In this context, the use of smart homes has been proposed as a means of improving access to home autonomy for people with ID by promoting their autonomy and control over their environment and by adapting to their lifestyle and their abilities (Lussier-Desrochers et al., 2008).

### 2.3 Adaptation and validation of a digital home assistance platform (Landuran et al., 2023)

Access to autonomous housing for people with disabilities is one of the issues related to inclusion and social participation (French law of 2005; Convention on the rights of people with disabilities, 2006) and constitutes a major step in personal development. Creating a "home" that meets people's needs has important implications for psychological well-being, community participation, daily autonomy and the emergence of self-determination (Fänge & Iwarsson, 2003).

However, studies show that people with ID do not fully participate in their lives and communities (Foley et al., 2013). Most adults live with their parents and not in self-contained accommodation (Foley et al., 2013). The lack of suitable housing (Bigby, 2010) and the cognitive and adaptation difficulties of this population can constitute an obstacle important in access to this residential autonomy.

As part of the work we present here (Landuran et al., 2023) we adapted the assisted living platform called DomAssist to the specificities of people with ID (Dupuy et al., 2016). It consists of sensors installed in the house (for example, motion sensors, contact sensors or electricity consumption sensors) and digital tablets allowing interactions between the user and the platform (Fig. 3.).

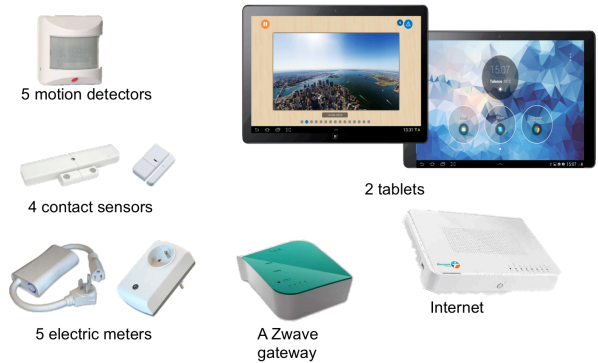

**Fig. 3.** The different elements of the home platform.

The support services offered by this platform cover three areas (Fig. 4.): security and safety (monitoring the front door and electrical appliances); assistance in carrying out daily activities (reminder of activities or appointments); assistance with social participation (simplified email, information on social events, etc.).

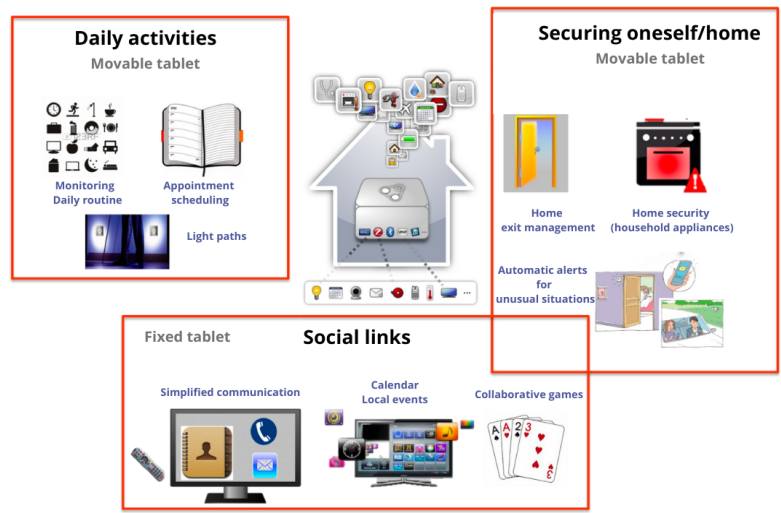

**Fig. 4.** Services offered by the DomAssist home assistance platform.

We assessed the impact of prolonged (six-month period) use of the platform on home living skills, self-determination, quality of life, self-esteem, level of worry, and well-being. Participants with ID were divided into two groups, the first equipped with the home assistance platform, the second (control group) playing games on a digital tablet.

The evaluations showed (Fig. 5.) a greater improvement on average for the experimental group compared to the control group for: home skills, self-determination, self-esteem, quality of life and psychological well-being in the dimensions of autonomy, skills and meaning in life. A significant reduction in the level of worry was also observed in the experimental group.

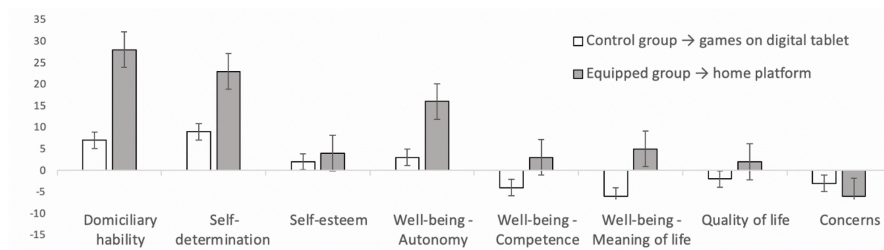

**Fig. 5.** Evolution of perceptions, in the different areas studied, after using the home assistance platform, for the two groups.

## 3  Autism Spectrum Disorder and self-determination

Autism Spectrum Disorder (ASD) is defined as a lifelong neurodevelopmental disorder characterized by two main symptoms: persistent deficits in social communication/interaction and restricted, repetitive behaviors (Bougeard et al., 2021). People with ASD may also have deficits intelligence and motor functions, as well as high levels of anxiety, stress, depression and isolation (Hudson et al., 2019; Salari et al., 2022). In addition to these various issues, much research has also focused on the level of self-determination and its consequences for people with ASD at different stages of their lives.

Some research has notified than children and youth with ASD have limited opportunities and supports to engage in self-determined actions in their environments (Moran et al., 2021) as well as lower levels of self-determination in comparison with their peers without disabilities (Shogren et al., 2018) and with other disabilities (Chou et al., 2017; Shogren et al., 2018). For example, Chou et al. (2017) showed that high school students with ASD had lower levels of autonomy and psychological empowerment than students with intellectual or learning disabilities. This lower level of self-determination is also linked to a lower level of quality of life and lower satisfaction with life (Shogren et al., 2008; White et al., 2018). Certain factors also appear to further reduce the level of self-determination, such as the severity of the disorder (White et al., 2022), young age (Wehmeyer et al., 2011) and male gender (Carter et al., 2013).

In adulthood, self-determination is also essential for people with ASD. Kim (2019) identified in a qualitative review that self-determined behaviors of adults with autism as goal-setting, decision making, problem solving, and self-management had positive influences on their employment status, social participation, positive identity, and stress management. Young adults with Autism Spectrum Disorder (ASD) also frequently encounter a reduced quality of life compared to their peers of similar age and abilities (Bishop-Fitzpatrick et al., 2018). Additionally, they often face challenges in attaining independence in their living arrangements (Steinhausen et al., 2016), and struggle to reach the typical developmental milestones associated with adulthood (Picci & Scherf, 2015). A large number of studies carried out with people with disabilities, including ASD, have also shown that the consequences of a low level of self-determination and independence can lead to poorer in-school and post-school outcomes such as advancing to higher education and gaining employment (Moran et al., 2021).

At university, students with ASD also face many challenges including difficulties with group work, peer relationships, ambiguous instructions, time management, navigating a lack of structure at university, etc (McPeake et al., 2023). These various academic difficulties can lead ASD students to experience fatigue, anxiety, depression, burn-out, dropping out of classes, suicidal ideation and attempting suicide, at higher levels than neurotypical students (Jackson et al., 2018; McPeake et al., 2023). They are also led to graduate at significantly lower rates when compared to their typically developing peers and students with other disabilities (Davis et al., 2021) And even if they manage to graduate, among all disabled graduates (at all qualification levels), graduates with autism are most likely to be unemployed (Allen & Coney, 2019).

In this context, supporting the self-determination of people with ASD is therefore a central issue, in particular by enabling them to access and succeed in higher education and the jobs of their choice.

### 3.1  Assistive technologies and self-determination in college students with ASD

Given the difficulties encountered by ASD students at university, various technologies are being developed to assist and help them in different areas of their life as university students.

Some technologies (e.g.: mobile app, software, learning management systems, etc) are developed to assist learning for students with ASD : 1) to reinforce the acquisition of certain concepts (McMahon et al., 2016; Begel et al., 2021; Dahlstrom-Hakki & Wallace, 2022); 2) to facilitate student engagement during the course session (Francis et al., 2018; Huffman et al., 2019) as well as outside by improving their revisions (Francis et al., 2018; O'Neill & Smyth, 2023); 3) to follow distance education (Satterfield et al., 2015; Richardson, 2017; Adams et al., 2019; Madaus et al., 2022). For example, Huffman et al, (2019) showed the efficacy of a self-monitoring application in university lectures to assist a student with ASD by allowing him to be more engaged and concentrated during the lesson.

Over technologies and digital tools are developed to assist or improve communication and interaction skills of students with ASD at university, mainly involving training in social interaction and communication via online course modules (PPT, videos, etc.), coaching (telecoaching, audioching) and video-modeling (Mason et al., 2012; Mason et al., 2020; Gregori et al., 2021), often combined together.

Other technologies are designed to enhance different cognitive and executive abilities, mainly on spatial navigation through mobile applications that can help students with ASD to get around more easily and to carry out activities, particularly on the university campus (McMahon et al., 2015; Kearney et al., 2020; Wright et al., 2020). For example, McMahon et al. (2015) showed the effectiveness of a location-based augmented reality application to teach one college student with ASD to navigate a city independently to find local employment opportunities.

Finally, Hrabal et al. (2023) showed that certain technologies (e.g.: video modeling), which are not specifically intended for college students, can help them improve daily living skills such as meal preparation and housekeeping tasks.

In general, these technology and digital tools appear to benefit university students with ASD, both in terms of effectiveness and satisfaction with their use, particularly for abilities essential to self-determination (learning, self-management, communication, spatial navigation, etc.). However, it is important to exercise caution when interpreting these findings for two principal reasons: 1) most of the studies involve very few participants (1 to 4) and are case studies; 2) there is considerable variability in the technologies and abilities targeted.

### 3.2 Atypie-Friendly and development of a virtual visit system for the University of Bordeaux

In France, Atypie-Friendly is a national program involving 25 universities, which aims to promote the inclusion and support of people with neurodevelopmental disorders, particularly ASD, at university. The Atypie-Friendly program addresses many aspects of students' lives, including the development of digital tools and projects to support the transition from high school to university.

Indeed, arriving at university is a crucial time for students. It involves radical changes compared with life at high school, with new teaching styles, new people, new facilities and so on. This transition can lead to difficulties in establishing new routines and can cause anxiety, particularly for students with ASD (Jill Boucher, 1977; van Steensel, 2011). These students may also experience difficulties with spatial navigation (Ring et al., 2018). A number of schemes have been put in place to overcome these difficulties. Universities generally organize open days to allow college students to discover the main services in their future environment. Individual detailed visits to their study site, to familiarize them with their future environment and their main contacts, can also be offered if required. However, it is difficult to envisage the possibility of carrying out visits for all new students. In addition, a number of factors (availability, geographical distance, etc.) mean that not all students who so wish can take part in these open days or individual visits.

Based on this premise, we are currently developing a virtual visit of key services at the University of Bordeaux (Health Service, Disability Service, Housing Service, University Restaurant). These visits will focus on 4 key themes necessary for the development of ASD student's self-determination (health and well-being, studies and disability, accommodation and catering).

Each visit will comprise (Fig. 6.): a "passive" guided digital visit in first person view of the key service (the person will be able to watch the journey scroll from a given point to the destination); an active visit (the person will be able to complete the journey themselves using mouse and keyboard); once at the destination, a presentation of the main tasks of the service by a resource person (passive mode: video; active mode: interactive video with clickable questions). The aim of setting up an active visit is to improve information retention, particularly spatial navigation (Cogné et al., 2016).

In order to test the effectiveness of these virtual visits, a protocol will be put in place to compare these three systems: 1) the current University system (information on the University website and use of Google Maps); 2) the passive virtual visit system; 3) the active virtual visit system. The effectiveness of these systems will be compared in terms of information retention (remembering the route, information about the service) and usability.

One of the project's future enhancements will be to integrate a conversational user interface including a conversational agent based on artificial intelligence into the virtual visit system. Conversational user interfaces, whether through text-based chat or voice recognition with synthesized responses, offer flexibility, personalization, and alternative communication methods to perform tasks for users (Iniesto et al., 2023). This conversational agent could complement the virtual visit system by accompanying students on their administrative processes (e.g.: assistance in disability disclosure and support) and in presenting in more details university services, resources and the administrative procedures necessary to access it. This point is extremely important to the extent that the complexity of administrative procedures constitutes a significant obstacle to the entry into university of students with ASD, and more generally of students with disabilities (Iniesto et al., 2023).

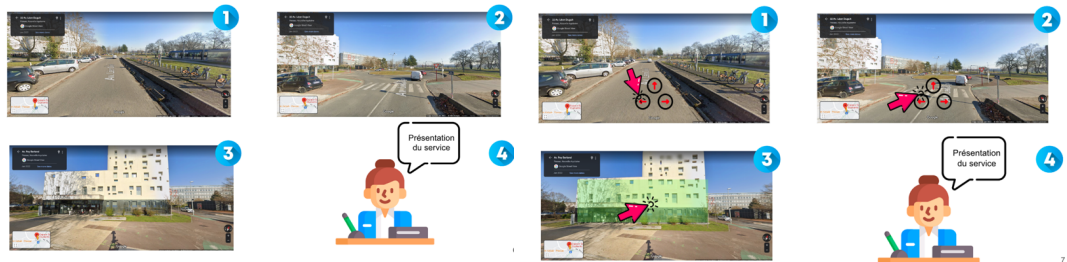

**Fig. 6.** Virtual visit to the Student Health Centre at the University of Bordeaux. Left: illustration of the passive virtual visit with 3 intersections. Right: illustration of the active virtual visit.

## 4      Conclusion

Autonomy and support for the self-determination of vulnerable people is one of the major societal challenges of the years to come.

Nirje (1972) was one of the first to use the term self-determination for people with disabilities in his chapter "The right to self-determination" which deals with the principle of normalization. This author clearly expressed the importance of personal self-determination for all, without excluding people with mental retardation or other disabilities. For him, a major challenge of the principle of normalization consists of creating conditions by which a person with disabilities experiences the respect and dignity to which every human being has the right. Thus, any action that concerns a person must take into account their choices, wishes, desires, preferences and aspirations.

In France, this idea is expressed through the law on "equal rights and opportunities, participation and citizenship of disabled people", adopted in 2005. In this law, the ability to express a life project is the starting point for procedures allowing disabled people to obtain compensation (financial, human, etc.), in order to help them realize their projects. But despite this impulse, too few people with disabilities live independently, are employed full time, hold paid employment or pursue postsecondary education (Steinhausen et al., 2016; Moran et al., 2021). Furthermore, large numbers of people with disabilities remain dependent on caregivers, service providers, and overburdened social systems (Wehmeyer, 1992), and have few opportunities to make choices based on their interests and abilities and find their lives controlled by other people who too often make decisions for them (Kozleski and Sands, 1992; Stancliffe, 1995).

In recent years, technologies supporting self-determination have proven to be extremely promising solutions. In this article we have seen that many assistive technologies are being tested in the scientific literature to assist people with ID and ASD in many areas of life (Näslund and Gardelli, 2013; McMahon et al., 2016). However, several limitations in access and use of digital technologies by these populations have been identified in the literature. They concern, in part, the lack of accessibility and the lack of technologies adapted to people's needs (Caton and Chapman, 2016). The authors also underline the importance of taking precautions given the great variability of the technologies tested and the small number of participants often included in the protocols (McMahon et al., 2015). To address these limitations, authors have proposed the use of universal design methodology (Wehmeyer et al., 2012), as well as different types of methodologies that largely include the user in the design cycle (Greenbaum and Kyng, 1991).

In our digital tool design activities, we combine different methods such as a user-centered design approach including participatory design, principles of universal design, recommendations from the literature and results from our empirical studies. Future users, families and professionals are placed at the center of the design process, through an iterative evaluation of the proposed solutions, at each stage of the process. This approach ensures the best possible usability of the tools developed

But in general, these digital tools are far from being adapted to all possible forms of user disability. In this area, technological progress and prospects for development are considerable in all areas of digital personal assistance (home automation, robotics, social communication, digital tablets, etc.). The need to adapt human-machine interfaces in order to promote the autonomy of different populations of people is an economic, societal and public health issue. Recent progress in Artificial Intelligence and the perspectives offered by intelligent systems or conversational agents, or even cobotics in cooperation with humans, open extremely promising issues in helping autonomy and self-determination for vulnerable people but also for everyone.

# 5 References

1. Adams, D., Simpson, K., Davies, L., Campbell, C., & Macdonald, L. (2019). Online learning for university students on the autism spectrum: A systematic review and questionnaire study. Australasian Journal of Educational Technology, 35(6), Article 6. https://doi.org/10.14742/ajet.5483
2. Alberto, P. A., Cihak, D. F., & Gama, R. I. (2005). Use of static picture prompts versus video modeling during simulation instruction. Research in Developmental Disabilities, 26(4), 327–339. https://doi.org/10.1016/j.ridd.2004.11.002
3. Allen, K. D., Burke, R. V., Howard, M. R., Wallace, D. P., & Bowen, S. L. (2012). Use of Audio Cuing to Expand Employment Opportunities for Adolescents with Autism Spectrum Disorders and Intellectual Disabilities. Journal of Autism and Developmental Disorders, 42(11), 2410–2419. https://doi.org/10.1007/s10803-012-1519-7
4. Allen, M., & Coney, K. (2019). What Happens Next? A Report on the First Destinations of 2017 Disabled Graduates. https://www.agcas.org.uk/write/MediaUploads/Resources/Disability%20TG/What_Happens_Next_report_2019.pdf.
5. Angyal, A. (1941). Foundations for a science of personality (pp. xii, 398). Commonwealth Fund.
6. Ayres, K., & Cihak, D. (2010). Computer- and Video-Based Instruction of Food-Preparation Skills: Acquisition, Generalization, and Maintenance. Intellectual and Developmental Disabilities, 48(3), 195–208. https://doi.org/10.1352/1944-7558-48.3.195
7. Ayres, K. M., Langone, J., Boon, R. T., & Norman, A. (2006). Computer-Based Instruction for Purchasing Skills. Education and Training in Developmental Disabilities, 41(3), 253–263.
8. Begel, A., Dominic, J., Phillis, C., Beeson, T., & Rodeghero, P. (2021). How a Remote Video Game Coding Camp Improved Autistic College Students' Self-Efficacy in Communication. Proceedings of the 52nd ACM Technical Symposium on Computer Science Education, 142–148. https://doi.org/10.1145/3408877.3432516
9. Bigby, C. (2010). A Five-Country Comparative Review of Accommodation Support Policies for Older People With Intellectual Disability. Journal of Policy and Practice in Intellectual Disabilities, 7(1), 3–15. https://doi.org/10.1111/j.1741-1130.2010.00242.x
10. Bishop-Fitzpatrick, L., Mazefsky, C. A., & Eack, S. M. (2018). The combined impact of social support and perceived stress on quality of life in adults with autism spectrum disorder and without intellectual disability. Autism, 22(6), 703–711. https://doi.org/10.1177/1362361317703090
11. Boucher, J. (1977). Alternation and Sequencing Behaviour, and Response to Novelty in Autistic Children. Journal of Child Psychology and Psychiatry, 18(1), 67–72. https://doi.org/10.1111/j.1469-7610.1977.tb00417.x
12. Bouck, E. C., Bassette, L., Taber-Doughty, T., Flanagan, S. M., & Szwed, K. (2009). Pentop Computers as Tools for Teaching Multiplication to Students with Mild Intellectual Disabilities. Education and Training in Developmental Disabilities, 44(3), 367–380.
13. Bougeard, C., Picarel-Blanchot, F., Schmid, R., Campbell, R., & Buitelaar, J. (2021). Prevalence of Autism Spectrum Disorder and Co-morbidities in Children and Adolescents: A Systematic Literature Review. Frontiers in Psychiatry, 12. https://www.frontiersin.org/articles/10.3389/fpsyt.2021.744709
14. Brandão, A., Trevisan, D. G., Brandão, L., Moreira, B., Nascimento, G., Vasconcelos, C. N., Clua, E., & Mourão, P. (2010). Semiotic Inspection of a Game for Children with Down Syndrome. 2010 Brazilian Symposium on Games and Digital Entertainment, 199–210. https://doi.org/10.1109/SBGAMES.2010.24

15. Brown, D. J., McHugh, D., Standen, P., Evett, L., Shopland, N., & Battersby, S. (2011). Designing location-based learning experiences for people with intellectual disabilities and additional sensory impairments. Computers & Education, 56(1), 11–20. https://doi.org/10.1016/j.compedu.2010.04.014
16. Bruni, M. (2006). Fine motor skills for children with Down syndrome (second edition). Bethesda, MD: Woodbine House
17. Cannella-Malone, H. I., Fleming, C., Chung, Y.-C., Wheeler, G. M., Basbagill, A. R., & Singh, A. H. (2011). Teaching Daily Living Skills to Seven Individuals With Severe Intellectual Disabilities: A Comparison of Video Prompting to Video Modeling. Journal of Positive Behavior Interventions, 13(3), 144–153. https://doi.org/10.1177/1098300710366593
18. Cannella-Malone, H., Sigafoos, J., O'Reilly, M., de la Cruz, B., Edrisinha, C., & Lancioni, G. E. (2006). Comparing Video Prompting to Video Modeling for Teaching Daily Living Skills to Six Adults with Developmental Disabilities. Education and Training in Developmental Disabilities, 41(4), 344–356.
19. Carter, E. W., Lane, K. L., Cooney, M., Weir, K., Moss, C. K., & Machalicek, W. (2013). Parent Assessments of Self-determination Importance and Performance for Students with Autism or Intellectual Disability. American Journal on Intellectual and Developmental Disabilities, 118(1), 16–31. https://doi.org/10.1352/1944-7558-118.1.16
20. Caton, S., & Chapman, M. (2016). The use of social media and people with intellectual disability: A systematic review and thematic analysis. Journal of Intellectual & Developmental Disability, 41(2), 125–139. https://doi.org/10.3109/13668250.2016.1153052
21. Chou, Y.-C., Wehmeyer, M. L., Palmer, S. B., & Lee, J. (2017). Comparisons of Self-Determination Among Students With Autism, Intellectual Disability, and Learning Disabilities: A Multivariate Analysis. Focus on Autism and Other Developmental Disabilities, 32(2), 124–132. https://doi.org/10.1177/1088357615625059
22. Cogné, M., Taillade, M., N'Kaoua, B., Tarruella, A., Klinger, E., Larrue, F., Sauzéon, H., Joseph, P.-A., & Sorita, E. (2017). The contribution of virtual reality to the diagnosis of spatial navigation disorders and to the study of the role of navigational aids: A systematic literature review. Annals of Physical and Rehabilitation Medicine, 60(3), 164–176. https://doi.org/10.1016/j.rehab.2015.12.004
23. Convention on the Rights of Persons with Disabilities (CRPD) | Division for Inclusive Social Development (DISD). (2006). Retrieved September 25, 2023, from https://social.desa.un.org/issues/disability/crpd/convention-on-the-rights-of-persons-with-disabilities-crpd
24. Cross, S., & Markus, H. (2010). Possible Selves across the Life Span. Human Development, 34(4), 230–255. https://doi.org/10.1159/000277058
25. Dahlstrom-Hakki, I., & Wallace, M. L. (2022). Teaching Statistics to Struggling Students: Lessons Learned from Students with LD, ADHD, and Autism. Journal of Statistics and Data Science Education, 30(2), 127–137. https://doi.org/10.1080/26939169.2022.2082601
26. D'Ateno, P., Mangiapanello, K., & Taylor, B. A. (2003). Using Video Modeling to Teach Complex Play Sequences to a Preschooler with Autism. Journal of Positive Behavior Interventions, 5(1), 5–11. https://doi.org/10.1177/10983007030050010801
27. Davies, D. K., Stock, S. E., Holloway, S., & Wehmeyer, M. L. (2010). Evaluating a GPS-Based Transportation Device to Support Independent Bus Travel by People With Intellectual Disability. Intellectual and Developmental Disabilities, 48(6), 454–463. https://doi.org/10.1352/1934-9556-48.6.454
28. Davies, D. K., Stock, S. E., & Wehmeyer, M. L. (2003). A Palmtop Computer-Based Intelligent Aid for Individuals with Intellectual Disabilities to Increase Independent Decision

Making. Research and Practice for Persons with Severe Disabilities, 28(4), 182–193. https://doi.org/10.2511/rpsd.28.4.182

29. Dupuy, L., Consel, C., & Sauzéon, H. (2016). Une assistance numérique pour les personnes âgées: Le projet DomAssist. https://inria.hal.science/hal-01278203
30. Fairweather, J. S., & Shaver, D. M. (1990). Making the Transition to Postsecondary Education and Training. Exceptional Children, 57(3), 264–270. https://doi.org/10.1177/001440299105700309
31. Fänge, A., & Iwarsson, S. (2003). Accessibility and usability in housing: Construct validity and implications for research and practice. Disability and Rehabilitation, 25(23), 1316–1325. https://doi.org/10.1080/09638280310001616286
32. Foley, K.-R., Jacoby, P., Girdler, S., Bourke, J., Pikora, T., Lennox, N., Einfeld, S., Llewellyn, G., Parmenter, T. R., & Leonard, H. (2013). Functioning and post-school transition outcomes for young people with Down syndrome. Child: Care, Health and Development, 39(6), 789–800. https://doi.org/10.1111/cch.12019
33. Francis, G. L., Duke, J. M., Kliethermes, A., Demetro, K., & Graff, H. (2018). Apps to Support a Successful Transition to College for Students With ASD. TEACHING Exceptional Children, 51(2), 111–124. https://doi.org/10.1177/0040059918802768
34. Gilson, C. B., & Carter, E. W. (2016). Promoting Social Interactions and Job Independence for College Students with Autism or Intellectual Disability: A Pilot Study. Journal of Autism and Developmental Disorders, 46(11), 3583–3596. https://doi.org/10.1007/s10803-016-2894-2
35. Greenbaum, J., & Kyng, M. (2020). Design at Work: Cooperative Design of Computer Systems. CRC Press.
36. Gregori, E., Mason, R., Wang, D., Griffin, Z., & Iriarte, A. (2022). Effects of Telecoaching on Conversation Skills for High School and College Students With Autism Spectrum Disorder. Journal of Special Education Technology, 37(2), 241–252. https://doi.org/10.1177/01626434211002151
37. Haro, B. P. M., Santana, P. C., & Magaña, M. A. (2012). Developing reading skills in children with Down syndrome through tangible interfaces. Proceedings of the 4th Mexican Conference on Human-Computer Interaction, 28–34. https://doi.org/10.1145/2382176.2382183
38. Hrabal, J. M., Davis, T. N., & Wicker, M. R. (2023). The Use of Technology to Teach Daily Living Skills for Adults with Autism: A Systematic Review. Advances in Neurodevelopmental Disorders, 7(3), 443–458. https://doi.org/10.1007/s41252-022-00255-9
39. Hudson, C. C., Hall, L., & Harkness, K. L. (2019). Prevalence of Depressive Disorders in Individuals with Autism Spectrum Disorder: A Meta-Analysis. Journal of Abnormal Child Psychology, 47(1), 165–175. https://doi.org/10.1007/s10802-018-0402-1
40. Huffman, J. M., Bross, L. A., Watson, E. K., Wills, H. P., & Mason, R. A. (2019). Preliminary Investigation of a Self-monitoring Application for a Postsecondary Student with Autism. Advances in Neurodevelopmental Disorders, 3(4), 423–433. https://doi.org/10.1007/s41252-019-00124-y
41. Iniesto, F., Coughlan, T., Lister, K., Devine, P., Freear, N., Greenwood, R., Holmes, W., Kenny, I., McLeod, K., & Tudor, R. (2023). Creating 'a Simple Conversation': Designing a Conversational User Interface to Improve the Experience of Accessing Support for Study. ACM Transactions on Accessible Computing, 16(1), 6:1-6:29. https://doi.org/10.1145/3568166
42. Jackson, S. L. J., Hart, L., Brown, J. T., & Volkmar, F. R. (2018). Brief Report: Self-Reported Academic, Social, and Mental Health Experiences of Post-Secondary Students with Autism Spectrum Disorder. Journal of Autism and Developmental Disorders, 48(3), 643–650. https://doi.org/10.1007/s10803-017-3315-x

43. Jarrold, C., & Baddeley, A. (2001). Short-term memory in Down syndrome: Applying the working memory model. Down Syndrome Research and Practice, 7(1), 17–23. https://doi.org/10.3104/reviews.110
44. Kagohara, D. M. (2011). Three Students with Developmental Disabilities Learn to Operate an iPod to Access Age-Appropriate Entertainment Videos. Journal of Behavioral Education, 20(1), 33–43. https://doi.org/10.1007/s10864-010-9115-4
45. Kagohara, D. M., van der Meer, L., Ramdoss, S., O'Reilly, M. F., Lancioni, G. E., Davis, T. N., Rispoli, M., Lang, R., Marschik, P. B., Sutherland, D., Green, V. A., & Sigafoos, J. (2013). Using iPods® and iPads® in teaching programs for individuals with developmental disabilities: A systematic review. Research in Developmental Disabilities, 34(1), 147–156. https://doi.org/10.1016/j.ridd.2012.07.027
46. Kearney, K. B., Joseph, B., Finnegan, L., & Wood, J. (2021). Using a Peer-Mediated Instructional Package to Teach College Students With Intellectual and Developmental Disabilities to Navigate an Inclusive University Campus. The Journal of Special Education, 55(1), 45–54. https://doi.org/10.1177/0022466920937469
47. Kim, S. Y. (2019). The experiences of adults with autism spectrum disorder: Self-determination and quality of life. Research in Autism Spectrum Disorders, 60, 1–15. https://doi.org/10.1016/j.rasd.2018.12.002
48. Kozleski, E. B., & Sands, D. J. (1992). The Yardstick of Social Validity: Evaluating Quality of Life as Perceived by Adults without Disabilities. Education and Training in Mental Retardation, 27(2), 119–131.
49. Lachapelle, Y., Lussier-Desrochers, D., Caouette, M., & Therrien-Bélec, M. (2013). Expérimentation d'une technologie mobile d'assistance à la réalisation de tâches pour soutenir l'autodétermination de personnes présentant une déficience intellectuelle. Revue francophone de la déficience intellectuelle, 24, 96–107. https://doi.org/10.7202/1021267ar
50. Lachapelle, Y., Wehmeyer, M. L., Haelewyck, M.-C., Courbois, Y., Keith, K. D., Schalock, R., Verdugo, M. A., & Walsh, P. N. (2005). The relationship between quality of life and self-determination: An international study. Journal of Intellectual Disability Research, 49(10), 740–744. https://doi.org/10.1111/j.1365-2788.2005.00743.x
51. Lancioni, G. E., O'Reilly, M. F., Dijkstra, A. W., Groeneweg, J., & Van den Hof, E. (2000). Frequent Versus Nonfrequent Verbal Prompts Delivered Unobtrusively: Their Impact on the Task Performance of Adults with Intellectual Disability. Education and Training in Mental Retardation and Developmental Disabilities, 35(4), 428–433.
52. Landuran, A., & N'Kaoua, B. (2018). Projet de vie: Regard des adultes avec une trisomie 21 et des aidants (familles et professionnels). Revue francophone de la déficience intellectuelle, 28, 61–69. https://doi.org/10.7202/1051099ar
53. Landuran, A., & N'Kaoua, B. (2021). Designing a Digital Assistant for Developing a Life Plan. International Journal of Human–Computer Interaction, 37(18), 1749–1759. https://doi.org/10.1080/10447318.2021.1908669
54. Landuran, A., Sauzéon, H., Consel, C., & N'Kaoua, B. (2023). Evaluation of a smart home platform for adults with Down syndrome. Assistive Technology, 35(4), 347–357. https://doi.org/10.1080/10400435.2022.2075487
55. Lanfranchi, S., Jerman, O., Dal Pont, E., Alberti, A., & Vianello, R. (2010). Executive function in adolescents with Down Syndrome. Journal of Intellectual Disability Research, 54(4), 308–319. https://doi.org/10.1111/j.1365-2788.2010.01262.x
56. LUSSIER-DESROCHERS, D., LACHAPELLE, Y., PIGOT, H., & BEAUCHET, J. (2008). Des habitats intelligents pour promouvoir l'autodétermination et l'inclusion sociale. Des Habitats Intelligents Pour Promouvoir l'autodétermination et l'inclusion Sociale, 18, 53–64.

57. LOI n° 2005-102 du 11 février 2005 pour l'égalité des droits et des chances, la participation et la citoyenneté des personnes handicapées et liens vers les décrets d'application—Dossiers législatifs—Légifrance. (2005). Retrieved September 25, 2023, from https://www.legifrance.gouv.fr/dossierlegislatif/JORFDOLE000017759074/
58. Madaus, J., Cascio, A., & Gelbar, N. (2022). Perceptions of College Students with Autism Spectrum Disorder on the Transition to Remote Learning During the COVID-19 Pandemic. Developmental Disabilities Network Journal, 2(2). https://digitalcommons.usu.edu/ddnj/vol2/iss2/5
59. Mason, R. A., Gregori, E., Wills, H. P., Kamps, D., & Huffman, J. (2020). Covert Audio Coaching to Increase Question Asking by Female College Students with Autism: Proof of Concept. Journal of Developmental and Physical Disabilities, 32(1), 75–91. https://doi.org/10.1007/s10882-019-09684-2
60. Mason, R. A., Rispoli, M., Ganz, J. B., Boles, M. B., & Orr, K. (2012). Effects of video modeling on communicative social skills of college students with asperger syndrome. Developmental Neurorehabilitation, 15(6), 425–434. https://doi.org/10.3109/17518423.2012.704530
61. McMahon, D., Cihak, D. F., & Wright, R. (2015). Augmented Reality as a Navigation Tool to Employment Opportunities for Postsecondary Education Students With Intellectual Disabilities and Autism. Journal of Research on Technology in Education, 47(3), 157–172. https://doi.org/10.1080/15391523.2015.1047698
62. McMahon, D. D., Cihak, D. F., Wright, R. E., & Bell, S. M. (2016). Augmented Reality for Teaching Science Vocabulary to Postsecondary Education Students With Intellectual Disabilities and Autism. Journal of Research on Technology in Education, 48(1), 38–56. https://doi.org/10.1080/15391523.2015.1103149
63. Mechling, L. C. (2008). Thirty Year Review of Safety Skill Instruction for Persons with Intellectual Disabilities. Education and Training in Developmental Disabilities, 43(3), 311–323.
64. Morán, M. L., Hagiwara, M., Raley, S. K., Alsaeed, A. H., Shogren, K. A., Qian, X., Gómez, L. E., & Alcedo, M. Á. (2021). Self-Determination of Students with Autism Spectrum Disorder: A Systematic Review. Journal of Developmental and Physical Disabilities, 33(6), 887–908. https://doi.org/10.1007/s10882-020-09779-1
65. Nair, K. P. S. (2003). Life goals: The concept and its relevance to rehabilitation. Clinical Rehabilitation, 17(2), 192–202. https://doi.org/10.1191/0269215503cr599oa
66. Näslund, R., & Gardelli, Å. (2013). 'I know, I can, I will try': Youths and adults with intellectual disabilities in Sweden using information and communication technology in their everyday life. Disability & Society, 28(1), 28–40. https://doi.org/10.1080/09687599.2012.695528
67. Nirje, B. (1972). The right to serf-determination. In W. Wolfensberger, Normalization: The principle of normalization. p. 176-200. Toronto: National Institute on Mental Retardation.
68. O'Neill, S. J., & Smyth, S. (2023). Using off-the-shelf solutions as assistive technology to support the self-management of academic tasks for autistic university students. Assistive Technology, 0(0), 1–15. https://doi.org/10.1080/10400435.2023.2230480
69. Picci, G., & Scherf, K. S. (2015). A Two-Hit Model of Autism: Adolescence as the Second Hit. Clinical Psychological Science, 3(3), 349–371. https://doi.org/10.1177/2167702614540646
70. Richardson, J. T. E. (2017). Academic attainment in students with autism spectrum disorders in distance education. Open Learning: The Journal of Open, Distance and e-Learning, 32(1), 81–91. https://doi.org/10.1080/02680513.2016.1272446

71. Ring, M., Gaigg, S. B., de Condappa, O., Wiener, J. M., & Bowler, D. M. (2018). Spatial navigation from same and different directions: The role of executive functions, memory and attention in adults with autism spectrum disorder. Autism Research, 11(5), 798–810. https://doi.org/10.1002/aur.1924
72. Rondal, J. A. (1995). Exceptional Language Development in Down Syndrome: Implications for the Cognition-Language Relationship. Cambridge University Press.
73. Salari, N., Rasoulpoor, S., Rasoulpoor, S., Shohaimi, S., Jafarpour, S., Abdoli, N., Khaledi-Paveh, B., & Mohammadi, M. (2022). The global prevalence of autism spectrum disorder: A comprehensive systematic review and meta-analysis. Italian Journal of Pediatrics, 48(1), 112. https://doi.org/10.1186/s13052-022-01310-w
74. Satterfield, D., Lepage, C., & Ladjahasan, N. (2015). Preferences for Online Course Delivery Methods in Higher Education for Students with Autism Spectrum Disorders. Procedia Manufacturing, 3, 3651–3656. https://doi.org/10.1016/j.promfg.2015.07.758
75. Shogren, K. A., Shaw, L. A., Raley, S. K., & Wehmeyer, M. L. (2018). Exploring the Effect of Disability, Race-Ethnicity, and Socioeconomic Status on Scores on the Self-Determination Inventory: Student Report. Exceptional Children, 85(1), 10–27. https://doi.org/10.1177/0014402918782150
76. Shogren, K. A., Wehmeyer, M. L., Palmer, S. B., Soukup, J. H., Little, T. D., Garner, N., & Lawrence, M. (2008). Understanding the Construct of Self-Determination: Examining the Relationship Between the Arc's Self-Determination Scale and the American Institutes for Research Self-Determination Scale. Assessment for Effective Intervention, 33(2), 94–107. https://doi.org/10.1177/1534508407311395
77. Sorrell, C. A., Bell, S. M., & McCallum, R. S. (2007). Reading Rate and Comprehension as a Function of Computerized versus Traditional Presentation Mode: A Preliminary Study. Journal of Special Education Technology, 22(1), 1–12. https://doi.org/10.1177/016264340702200101
78. Soto, G., Belfiore, P. J., Schlosser, R. W., & Haynes, C. (1993). Teaching Specific Requests: A Comparative Analysis on Skill Acquisition and Preference Using Two Augmentative and Alternative Communication Aids. Education and Training in Mental Retardation, 28(2), 169–178.
79. Stancliffe, R. J. (1995). Assessing opportunities for choice-making: A comparison of self- and staff reports. American Journal of Mental Retardation, 99(4), 418–429.
80. Steinhausen, H.-C., Mohr Jensen, C., & Lauritsen, M. B. (2016). A systematic review and meta-analysis of the long-term overall outcome of autism spectrum disorders in adolescence and adulthood. Acta Psychiatrica Scandinavica, 133(6), 445–452. https://doi.org/10.1111/acps.12559
81. Stock, S. E., Davies, D. K., Davies, K. R., & Wehmeyer, M. L. (2006). Evaluation of an application for making palmtop computers accessible to individuals with intellectual disabilities. Journal of Intellectual & Developmental Disability, 31(1), 39–46. https://doi.org/10.1080/13668250500488645
82. Taber-Doughty, T., Patton, S. E., & Brennan, S. (2008). Simultaneous and Delayed Video Modeling: An Examination of System Effectiveness and Student Preferences. Journal of Special Education Technology, 23(1), 1–18. https://doi.org/10.1177/016264340802300101
83. Van Gameren-Oosterom, H. B. M., Fekkes, M., Reijneveld, S. A., Oudesluys-Murphy, A. M., Verkerk, P. H., Van Wouwe, J. P., & Buitendijk, S. E. (2013). Practical and social skills of 16–19-year-olds with Down syndrome: Independence still far away. Research in Developmental Disabilities, 34(12), 4599–4607. https://doi.org/10.1016/j.ridd.2013.09.041

84. van Steensel, F. J. A., Bögels, S. M., & Perrin, S. (2011). Anxiety Disorders in Children and Adolescents with Autistic Spectrum Disorders: A Meta-Analysis. Clinical Child and Family Psychology Review, 14(3), 302–317. https://doi.org/10.1007/s10567-011-0097-0
85. Wehmeyer, M. L. (1992). Self-Determination and the Education of Students with Mental Retardation. Education and Training in Mental Retardation, 27(4), 302–314.
86. Wehmeyer, M. L. (1999). A Functional Model of Self-Determination: Describing Development and Implementing Instruction. Focus on Autism and Other Developmental Disabilities, 14(1), 53–61. https://doi.org/10.1177/108835769901400107
87. Wehmeyer, M. L. (2005). Self-Determination and Individuals with Severe Disabilities: Re-Examining Meanings and Misinterpretations. Research and Practice for Persons with Severe Disabilities, 30(3), 113–120. https://doi.org/10.2511/rpsd.30.3.113
88. Wehmeyer, M. L. (2013). The Oxford Handbook of Positive Psychology and Disability. OUP USA.
89. Wehmeyer, M. L., Abery, B. H., Zhang, D., Ward, K., Willis, D., Hossain, W. A., Balcazar, F., Ball, A., Bacon, A., Calkins, C., Heller, T., Goode, T., Dias, R., Jesien, G. S., McVeigh, T., Nygren, M. A., Palmer, S. B., & Walker, H. M. (2011). Personal Self-Determination and Moderating Variables that Impact Efforts to Promote Self-Determination. Exceptionality, 19(1), 19–30. https://doi.org/10.1080/09362835.2011.537225
90. Wehmeyer, M. L., Palmer, S. B., Lee, Y., Williams-Diehm, K., & Shogren, K. (2011). A Randomized-Trial Evaluation of the Effect of Whose Future Is It Anyway? On Self-Determination. Career Development for Exceptional Individuals, 34(1), 45–56. https://doi.org/10.1177/0885728810383559
91. Wehmeyer, M., & Schwartz, M. (1997). Self-Determination and Positive Adult Outcomes: A Follow-up Study of Youth with Mental Retardation or Learning Disabilities. Exceptional Children, 63(2), 245–255. https://doi.org/10.1177/001440299706300207
92. Wert, B. Y., & Neisworth, J. T. (2003). Effects of Video Self-Modeling on Spontaneous Requesting in Children with Autism. Journal of Positive Behavior Interventions, 5(1), 30–34. https://doi.org/10.1177/10983007030050010501
93. White, K., Flanagan, T. D., & Nadig, A. (2018). Examining the Relationship Between Self-Determination and Quality of Life in Young Adults with Autism Spectrum Disorder. Journal of Developmental and Physical Disabilities, 30(6), 735–754. https://doi.org/10.1007/s10882-018-9616-y
94. White, S. W., Smith, I., & Brewe, A. M. (2022). Brief Report: The Influence of Autism Severity and Depression on Self-Determination Among Young Adults with Autism Spectrum Disorder. Journal of Autism and Developmental Disorders, 52(6), 2825–2830. https://doi.org/10.1007/s10803-021-05145-y
95. Wright, R. E., McMahon, D. D., Cihak, D. F., & Hirschfelder, K. (2022). Smartwatch Executive Function Supports for Students With ID and ASD. Journal of Special Education Technology, 37(1), 63–73. https://doi.org/10.1177/0162643420950027
96. Wu, P.-F., Cannella-Malone, H. I., Wheaton, J. E., & Tullis, C. A. (2016). Using Video Prompting With Different Fading Procedures to Teach Daily Living Skills: A Preliminary Examination. Focus on Autism and Other Developmental Disabilities, 31(2), 129–139. https://doi.org/10.1177/1088357614533594
97. Wuang, Y.-P., Chiang, C.-S., Su, C.-Y., & Wang, C.-C. (2011). Effectiveness of virtual reality using Wii gaming technology in children with Down syndrome. Research in Developmental Disabilities, 32(1), 312–321. https://doi.org/10.1016/j.ridd.2010.10.002